\documentclass[12pt]{iopart}
\usepackage[dvips]{graphicx}
\begin{document}
\title{Exchange bias effect and intragranular magnetoresistance in Nd$_{0.84}$Sr$_{0.16}$CoO$_3$}
\author{M. Patra, S. Majumdar and S. Giri}
\ead{sspsg2@iacs.res.in} 
\address{Department of Solid State Physics and Center for Advanced Materials,Indian Association for the Cultivation of Science, Jadavpur, Kolkata 700 032, INDIA}
\begin{abstract}
Electrical transport properties as a function of magnetic field and time  have been investigated in polycrystalline, Nd$_{0.84}$Sr$_{0.16}$CoO$_3$. A strong exchange bias (EB) effect is observed associated with the fairly large intragranular magnetoresistance (MR). The EB effect observed in the MR curve is compared with the EB effect manifested in magnetic hysteresis loop. Training effect, described as the decrease of EB effect when the sample is successively field-cycled at a particular temperature, has been observed in the shift of the MR curve. Training effect could be analysed by the successful models. The EB effect, MR and a considerable time dependence in MR are attributed to the intrinsic nanostructure giving rise to the varieties of magnetic interfaces in the grain interior. 
\end{abstract}
\maketitle

\section{Introduction}
Since the evidence of spin-polarized electron transport \cite{john} followed by the discovery of giant magnetoresistance (GMR) in layered magnetic multilayer films \cite{baib,bina}, a twist in conventional electronics has been experienced and a new era in electronics begins, which is termed as spintronics. Spintronics or spin electronics is a multidisciplinary field where manipulation of spin degrees of freedom has been exploited in solid state electronics \cite{prinz,gregg}. Spintronics involving GMR, tunnel  magnetoresistance (TMR), and exchange bias (EB) effects has attracted a considerable attention in the last decade. In particular, a new generation of spintronic devices have been designed such as magnetoresistive random access memory and spin valve sensors. 

Exchange bias phenomenon was initially reported in a heterogeneous system composed of ferromagnetic (FM) and  antiferromagnetic (AFM) substances manifested by the shifts in the magnetic hysteresis loop when the system is cooled through the N\'{e}el temperature  \cite{meik}. Since the discovery of exchange bias effect it has been noticed in varieties of combination between soft and hard magnetic substances e.g. FM, AFM, spin-glass (SG), cluster-glass (CG), ferrimagnetic (FI) systems \cite{nogues,iglesias,patra2,patra3}. The EB  effect has also been evidenced through the systematic shifts in  magnetoresistance-field (MR-$H$) curve which is limited to the bilayer or multilayer films only   \cite{niko,kerr,dai,bea}. Recently, EB effect has been reported in few compounds attributed to the grain interior spontaneous phase separation where EB effect was reported  through the shift in the  magnetic hysteresis loop  \cite{patra2,patra3,patra1,patra1a,neib,tang1,tang2} 
In this article, we report a new paradigm of EB effect in a polycrystalline compound, Nd$_{0.84}$Sr$_{0.16}$CoO$_3$ through the shifts in the MR-$H$ curve. We observe the strong EB effect involved with a fairly large intragranular MR where the value of EB field is found to be much larger than the observation measured through the shift in the magnetic hysteresis loop. Furthermore, we observe training effect in the shift of the MR-$H$ curve which is in accordance with the successful models used to interpret the training effect observed in the magnetic hysteresis loop. A strong time dependence in the resistivity is observed at low temperature which is found to be correlated with the EB effect. Finally, we proposed a possible scenario of grain interior magnetic  nanostructure to interpret the new magnetotransport behavior involved with the EB effect.

The hole doped compound, Nd$_{1-x}$Sr$_x$CoO$_3$ exhibits different characteristic features depending on the degree of hole doping \cite{stau}. For the low doping range the SG or CG state has been proposed with resistivity showing a semiconducting temperature dependence. With further increase in hole doping the short range FM clusters begin to coalesce above a percolation threshold ($x >$ 0.18) to attain magnetic long range ordering and it shows metallic conductivity in the ordered state.
The coexistence of FI and FM ordering is reported for the compounds above and close to the percolation threshold. Neutron powder diffraction studies on Nd$_{0.67}$Sr$_{0.33}$CoO$_3$ confirm the coexistence of FM and FI ordering where ferrimagnetism was interpreted in terms of an induced antiparallel ordering of the Nd spins in close proximity to the Co sublattice \cite{krim}. In the present investigation we note that the convincing feature of EB effect in the MR-$H$ curve is observed at $x$ = 0.16 which is close to the percolation threshold.   

\begin{figure}[t]
\centering
\includegraphics[width = 7.5 cm]{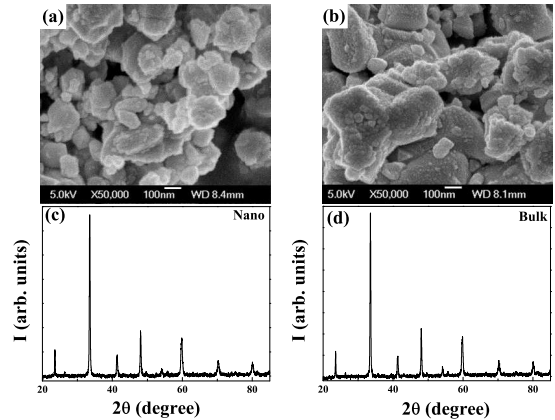}
\caption {SEM images for the nanoparticle (a) and bulk (b) with the corresponding powder x-ray diffraction patterns are shown in (c) and (d), respectively.}
\label{Fig. 1}
\end{figure}

\section{Synthesis and structural characterization}
Polycrystalline compound with composition Nd$_{0.84}$Sr$_{0.16}$CoO$_3$ was prepared by sol-gel technique \cite{patra2}. The preheated samples were finally annealed at 1073 and 1273 K with their average grain sizes $\sim$ 100 and $\sim$ 200 nm, respectively which were confirmed by Scanning Electron Microscopy using a JEOL FESEM microscope (JSM-6700F). The grain size of the samples are shown in figures 1(a) and 1(b). For simplicity, we address the sample with larger grain size as bulk while the sample with smaller grain size is defined as nanoparticle. Single phase of the orthorhombic structure ($Pbnm$) was confirmed for both the samples by a powder x-ray diffractometer (Seifert XRD 3000P) using CuK$_{\alpha}$ radiation. Figures 1(c) and 1(d) exhibit the x-ray diffraction patterns for the nanoparticle and bulk, respectively. In order to check any microstructural inhomogeneities in the grain interior, high resolution Transmission Electron Microscopy (HRTEM) was carried out on the bulk using a JEOL TEM, 2010 microscope. We have checked several portions of different particles randomly which do not show any structural inhomogeneities. An example of HRTEM image is shown in the left panel of figure 2 exhibiting a single  lattice fringe. The lattice spacing of the observed plane is $\sim$ 0.212 nm which matches with the spacing ($\sim$ 0.218 nm) of (202) plane of the orthorhombic structure ($Pbnm$) having lattice constants, $a$ = 0.534 nm, $b$ = 0.533 nm, and $c$ = 0.757 nm obtained from  the powder x-ray diffraction patterns. Two dimensional fast fourier transformation (FFT) of the lattice resolved image is shown in the right panel of figure 2 which can be indexed to the orthorhombic structure. Magnetic field dependence of resistivity ($\rho$) and magnetization ($M$) were carried out on the pelletised samples using a cryogen-free  physical property measurement system (Cryogenic Ltd., UK). 

\begin{figure}[t]
\centering
\includegraphics[width = 14.0 cm]{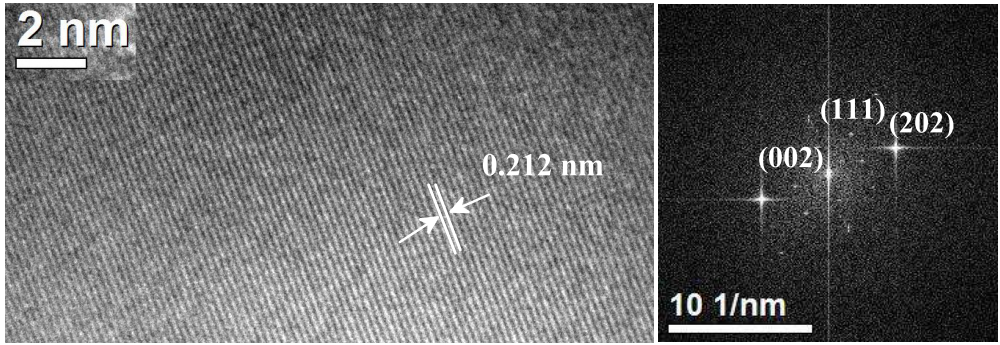}
\caption {HRTEM image (left panel) and corresponding FFT image (right panel) are shown for the bulk sample.}
\label{Fig. 2}
\end{figure}

\section{Results and discussions}
\subsection{Intragranular magnetoresistance and exchange bias effect}
Magnetoresistance (MR) is defined as ($\rho_{\rm H}$ - $\rho_0$)/$\rho_0$ where $\rho_{\rm H}$ and $\rho_0$ are the resistivities in a static and zero field, respectively. A symmetric MR-$H$ curve (open symbol) at 5 K having a fairly large MR ($\sim$ 12 \%) at 50 kOe is shown in figure 3 for the bulk. MR-$H$ curve shows a typical manifestation of tunneling mechanism exhibiting a peak in MR-$H$ curve. Tunneling between ferromagnetic (FM) grains accross the grain boundary region has been commonly interpreted in polycrystalline materials \cite{coey}. Since grain boundary region is considerably enhanced by decreasing the grain size than its bulk counterpart, MR should increase considerably due to the decrease in average grain size if grain boundary effect takes part dominating role in the magnetoresistance. MR-$H$ curve at 5 K was also measured for the  nanoparticle depicted by the closed symbols in figure 3. For the nanoparticle MR decreases considerably, suggesting that the grain boundary effect does not contribute any significant role. Rather, grain interior mechanism causes the change in MR. The short range FM and metallic  clusters embedded in the non-FM  matrix in the grain interior has been proposed by Stauffer {\it et al}. in Nd$_{1-x}$Sr$_x$CoO$_3$ where density and size of the short range FM clusters increase with hole (Sr) doping \cite{stau}. Thus, we propose that MR is involved with the tunneling between metallic FM clusters across the non-FM matrix analogous to that observed in the identical spontaneously phase separated cobaltite, La$_{0.85}$Sr$_{0.15}$CoO$_3$ \cite{wu2}. The grain interior nanostructure changes due to the change in average grain size which leads to the modification in the tunneling magnetoresistance. In addition to TMR, an asymmetry in the MR-$H$ curve is strikingly observed along with the convincing shifts in the vertical and horizontal axes (top panel of figure 4) when sample was cooled down to 5 K from room temperature in a cooling field, $H_{\rm cool}$ = 20 kOe. We further note that shifts are present in magnetic  hysteresis loop when sample was cooled under similar condition. MR and $(M/M_{50})^2$ with $H$ are plotted together in the bottom panel of figure 4 where $M_{50}$ is the magnetization ($M$) at $H$ = 50 kOe. Asymmetry in the MR-$H$ and $(M/M_{50})^2$-$H$ curves are the typical manifestation of EB effect. The measurements were carried out at $x$ = 0.12, 0.16, and 0.18 where EB effect is found to be largest at $x$ = 0.16 which is close to the percolation threshold of conductivity as well as long range FM ordering.   

\begin{figure}[t]
\centering
\includegraphics[width = 8 cm]{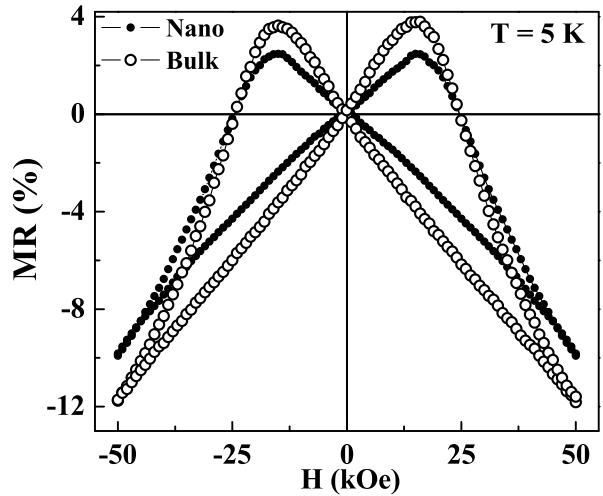}
\caption {MR-$H$ curves for  bulk and nanoparticles at 5 K while the sample was cooled down to 5 K in zero field.}
\label{Fig. 3}
\end{figure}

The peaks are typically observed in MR-$H$ curve at $M$ = 0 for TMR where magnetic field at the peak position determines the coercivity \cite{coey}. Here, the average value of the field at the peaks in MR is observed at $H_{\rm C}^\prime$ = 15.5 kOe which is almost twice of the coercivity ($H_{\rm C}$ = 8.0 kOe) obtained from the magnetic hysteresis loop. The peaks in the MR-$H$ and $(M/M_{50})^2$-$H$ curves, from which coercivities are estimated, are illustrated by the arrows at b$^{\prime}$, d$^{\prime}$, b, and d in the bottom panel of figure 4. Magnetoelectric phase diagram \cite{stau} and neutron diffraction \cite{krim} confirm that short range FM clusters are embedded in the ferrimagnetic (FI) matrix close to the percolation limit. If the system consists of FM and FI substances, bulk magnetic measurements provide the average coercivity of FM and FI components. It is recognized that coercivity of FM compound is typically much smaller than the anisotropic FI compound. Here, TMR is involved with the tunneling between metallic FM clusters across the semiconducting FI matrix where tunneling barrier is set by the anisotropy of the FI spins. Thus, coercivity noticed in MR-$H$ curve provides the coercivity of individual FI component. 

\begin{figure}[t]
\centering
\includegraphics[width = 11 cm]{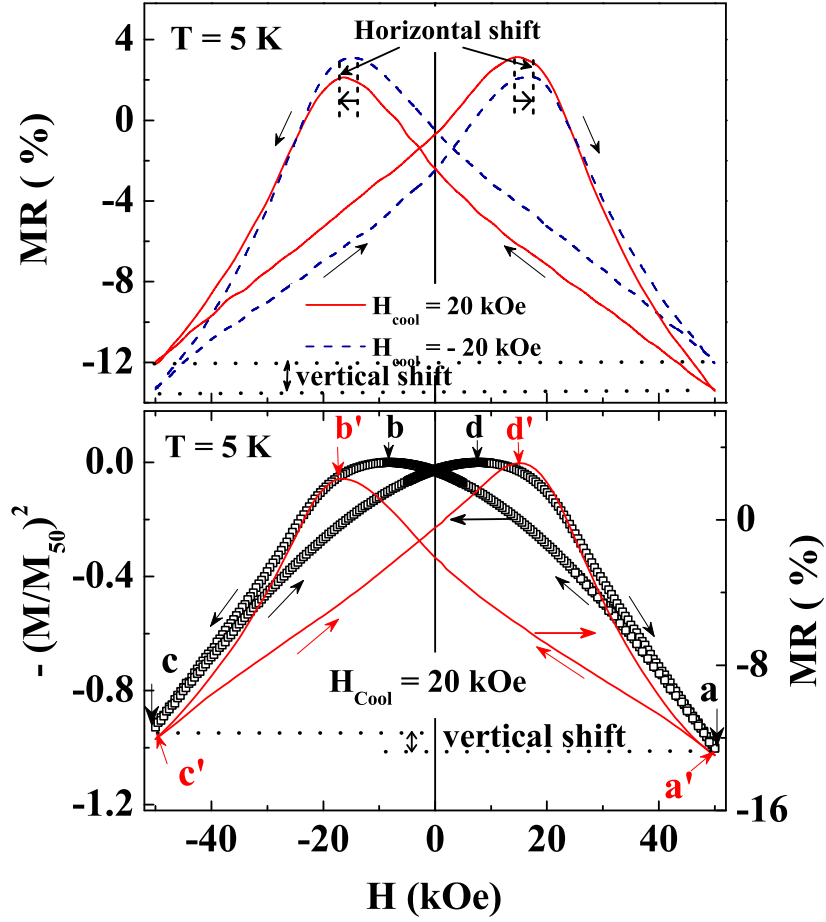}
\caption {Top panel: MR-$H$ curves for $H_{\rm cool}$ = $\pm$ 20 kOe displaying the vertical and horizontal shifts. Bottom panel: $-(M/M_{50})^2$-$H$ and MR-$H$ curves with $H_{\rm cool}$ = 20 kOe for comparing the magnetic and magnetoresistance results.}
\label{Fig. 4}
\end{figure}

If $H_{\rm b}$ and $H_{\rm d}$ are the negative and positive coercivities indicated by the arrows at peaks b and d, respectively, EB field ($H_{\rm E}$) is defined as $H_{\rm E}$ = $\left|H_{\rm b} - H_{\rm d}\right|$/2 $\approx$ 250 Oe. Exchange bias field ($H_{\rm  E}^{\rm MR}$) obtained from the MR-$H$ curve is $H_{\rm E}^{\rm MR}$ = $\left|H_{\rm b^\prime}^{\rm MR} -  H_{\rm d^\prime}^{\rm MR}\right|$/2 $\approx$ 744 Oe where $H_{\rm b^\prime}^{\rm MR}$ and $H_{\rm d^\prime}^{\rm MR}$ correspond to the fields at peaks   b$^\prime$ and d$^\prime$, respectively depicted in the bottom panel of figure 4.  Interestingly, $H_{\rm E}^{\rm MR}$ is significantly much higher than $H_{\rm E}$. Theoretical interpretations  \cite{meik1,wein} seem to agree with the conclusion that EB effect should be stronger for larger anisotropy of the hard magnetic substance. Here, $H_{\rm E}^{\rm MR} >> H_{\rm E}$ is in accordance with the proposed theories \cite{meik1,wein}. 
We note in the top panel of figure 4 that the vertical and horizontal shifts in MR are opposite while the direction of $H_{\rm cool}$ is opposite. The peak position in MR-$H$ curve measured from 50 kOe to -50 kOe is shifted toward negative direction along with the considerable decrease in the height of the curve while the curve measured from -50 kOe to 50 kOe is remained almost unaltered for $H_{\rm cool}$ = 20 kOe. The opposite feature is illustrated in the top panel of figure 4 for $H_{\rm cool}$ = -20 kOe. The results clearly demonstrate the spin valve-like character in MR analogous to that  reported in bilayer or multilayer films \cite{niko,bea,dai}.

\begin{figure}[t]
\centering
\includegraphics[width = 9 cm]{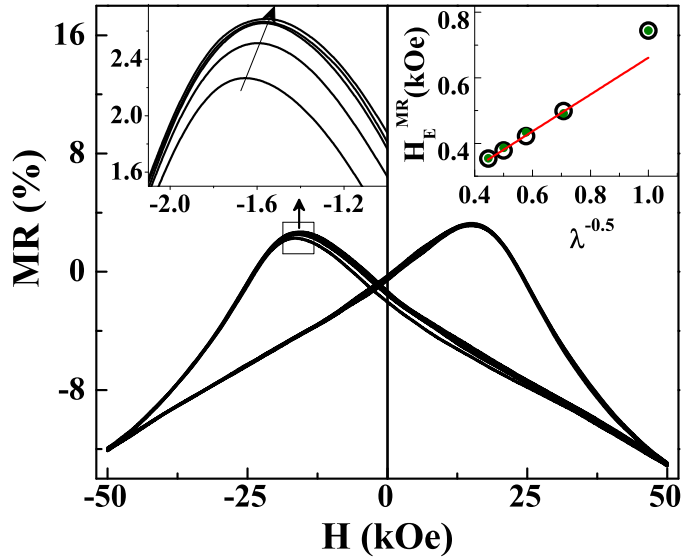}
\caption {Training effect is shown in MR-$H$ curves up to 5 successive cycles ($\lambda$ = 5). Left inset highlight the shift of peak position with increasing $\lambda$ where arrow indicates the increasing direction of $\lambda$. (Right inset) Plot of $H_{\rm E}$ with $\lambda^{-0.5}$ where solid straight line exhibits the fit by a power law. Open circles are the experimental data while small filled symbols represent the simulated data described in Eq. (1).}
\label{Fig. 5}
\end{figure}

\subsection{Training effect in the magnetoresistance curve}

Training effect (TE) is one of the significant manifestation of EB effect which describes the decrease of EB effect when the sample is successively field cycled at a particular temperature. TE is typically observed in the magnetic hysteresis loop \cite{nogues} which is recently reported in the MR-$H$ curve of only two multilayer films, exhibiting EB effect \cite{ven,brems}. In figure 5 typical signature of TE is illustrated at 5 K up to 5 successive cycles ($\lambda$). The decrease of the shift of peak positions along the negative field axis is highlighted in the left inset of the figure. A large decrease of $H_{\rm E}^{\rm MR}$ $\sim$ 33 \% is observed in between first and second cycles. The decrease of $H_{\rm E}^{\rm MR}$ is fitted satisfactorily (solid straight line) with the empirical relation, $H_{\rm E}^{\rm MR}(\lambda) - H_{\rm E}^{\rm MR}(\lambda = \infty) \propto \frac{1}{\sqrt{\lambda}}$ for $\lambda \geq$ 2 with $H_{\rm E}^{\rm MR}(\lambda = \infty)  \approx$ 100 Oe. The empirical relation does not fit the sharp decrease between first and second cycles in accordance with the reported results in the magnetic hysteresis loop \cite{nogues} as well as MR-$H$ curve  \cite{ven}. Binek proposed a recursive formula in the framework of spin configurational relaxation model to interpret training effect which correlates the ($\lambda$+1)th loop shift with the $\lambda$th one as \cite{binek}
\begin{equation}
H_{\rm E}^{\rm MR} (\lambda + 1) - H_{\rm E}^{\rm MR} (\lambda) = -\gamma [H_{\rm E}^{\rm MR} (\lambda) - H_{\rm E}^{\rm MR^\prime}(\lambda = \infty)]^3
\end{equation} 
where $\gamma$ is a sample dependent constant. Using $\gamma$ = 8.3 $\times$ 10$^{-7}$ Oe$^{-2}$ and $H_{\rm E}^{\rm MR^\prime}(\lambda = \infty)$ = 71.87 Oe the whole set of data (filled circles) could be generated which matches satisfactorily with the experimental data (open circles). 

\subsection{Time dependence in the resistivity}

We note a considerable time ($t$) dependence in $\rho$ at low temperature when sample was cooled down to 5 K from room temperature with $H_{\rm cool}$ = 50 kOe and then $\rho$ was measured with $t$ after removal of magnetic field. Plot of [$\rho$($t$) - $\rho$(0)]/$\rho$(0) with $t$ is illustrated in figure 6 (open circles). The relaxation follows the stretched exponential with a critical exponent ($\beta$). The value of $\beta$ is 0 $< \beta <$ 1, when it is involved with the activation against distribution of anisotropy barriers typically observed in the relaxation of magnetization, exhibiting glassy magnetic behavior \cite{mydosh}. The satisfactory fit using stretched exponential function, [$\rho$($t$) - $\rho$(0)]/$\rho$(0) = $A + B\exp(t/\tau)^{\beta}$ is shown by the continuous curve in the figure with relaxation time, $\tau$ = 4374 s and $\beta$ = 0.33, suggesting the glassy magnetic behavior in the transport properties. A very small $t$ dependence of $\rho$ is observed (filled circles) when sample was cooled in zero field and measurement was carried out in 50 kOe. The results clearly 
\begin{figure}[t]
\centering
\includegraphics[width = 11 cm]{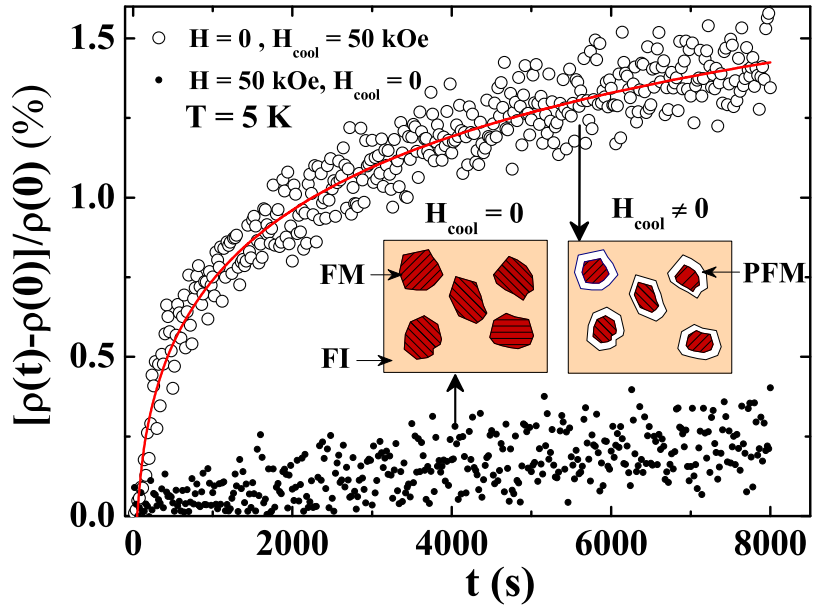}
\caption {Time ($t$) evolution of [$\rho$($t$) - $\rho$(0)]/$\rho$(0) at 5 K measured in zero field (open symbols) and $H$ = 50 kOe (filled symbols) after cooling in FC and ZFC modes, respectively. Inset shows cartoons of grain interior magnetic nanostructure. FM, PFM  and FI represent the FM, pinned FM and FI regions.}
\label{Fig. 6}
\end{figure}
demonstrate that strong $t$ evolution of $\rho$ is involved with field cooling which is correlated to the EB effect. HRTEM image confirms the absence of any  structural inhomogeneities in the grain interior of the compound, which indicates that grain interior nanostructure is purely magnetic in origin in accordance with the previous reports \cite{stau,krim}. A cartoon of the phase separation scenario within a grain is proposed in the left inset of figure 6 where short range FM regions or clusters are embedded in the FI matrix. $\rho(t)$ measured in two different conditions is illustrated in the figure while vertical arrows represent the corresponding grain interior magnetic nanostructure. When the sample is cooled in field, a new layer consisting of pinned ferromagnetic (PFM) spins arises at the FM/FI interface due to the pinning of the soft FM spins by the hard FI spins. The appearance of new layer comprising of PFM spins leads to the strong $t$ dependence in $\rho$, exhibiting glassy magnetic behavior in the transport properties. Furthermore, the PFM layer having unidirectional anisotropy gives rise to the unidirectional shift in the MR-$H$ curve where polarization direction of PFM spins are strongly influenced by the direction of cooling field leading to the EB effect and spin valve-like mechanism in MR.

\section{Conclusions}

In this article we report a new observation of EB effect in a compound attributed to the grain interior intrinsic nanostructure where EB is observed through the measurements of magnetoresistance analogous to that observed in the bilayer or multilayer films.
Training effect in the shift of the magnetoresistance curve further confirms the evidence of EB effect where training effect could be analyzed with the successful models. The EB effect is found to be substantially larger than the effect observed in the magnetic hysteresis loop where stronger EB effect is correlated to the stronger coercivity in MR than the magnetic measurements. A strong time dependence in the resistivity exhibiting glassy magnetic behavior is observed due to the field cooling which is correlated to the EB effect. The possible scenario of the grain interior nanostructure has been suggested to interpret the experimental results in polycrystalline, Nd$_{0.84}$Sr$_{0.16}$CoO$_3$ where grain interior nanostructure is close to the percolation threshold.

\vspace{0.3cm}
\noindent
{\bf Acknowledgements}\\
S.G. wishes to thank DST (Project No. SR/S2/CMP-46/2003), India for the financial support. M.P. thanks CSIR, India for the fellowship.

\end{document}